# Quantifying Electronic and Vibronic Contributions to Charge Transfer at the Nanoscale

Jessica Martinez[1], Eric Duverger[2], Michele Pavanello[1] and Damien Riedel[3]

[1]*Department of Chemistry, Rutgers University, Newark, New Jersey 07102, United States, USA.*

[2]*Institut FEMTO-ST, Université Marie et Louis Pasteur, CNRS, 15B avenue des Montboucons, F-25030 Besançon, France*

[3]*Institut des Sciences Moléculaires d'Orsay (ISMO), CNRS, Univ. Paris Sud, Université Paris-Saclay, F-91405 Orsay, France.*

**Abstract:**

Charge transfer (CT) is governed by complex multiscale dynamics sensitive to environmental factors[1,2,3]. In molecules, charge state and vibrational effects shape energy levels, charge distribution, and reactivity, thereby controlling CT efficiency[4,5,6,7]. Quantifying these contributions in CT is experimentally challenging, as vibrational effects remain difficult to isolate due to limits in precision, sensitivity, and stability. In such context, the original Marcus theory[8,9] often lacks the refinement required to accurately capture CT rates in complex environments, necessitating new approaches that incorporate vibronic effects[10]. Here, we examine a non-covalent $(H_2TPP)_2$ dyad on a semi-insulating $CaF_2$/Si(100) surface. Using a low-temperature (9 K) scanning tunneling microscope (STM), we generate tunneling electrons that trigger CT events by creating locally cationic states while activating transient vibronic modes at specific molecular sites in one fragment. This initiates a meso-N tautomerization serving as a CT signature in the second fragment. By tuning the tunnel electron energy, CT rates measurements reveal a periodic modulation. Quantitative analysis with the Marcus-Levich-Jortner model identifies key reorganization energies and resonant vibronic modes, while DFT provides complementary conformational and vibrational insights. Extending the Marcus-Antoniewicz models to explicit reaction coordinates reveals that CT is governed by the interplay between electronic and vibronic contributions, establishing surface-supported systems as a model framework.

CT processes underpin a wide range of physical, chemical, and biological phenomena, as well as key technologies including optoelectronics, nanoelectronics, and energy conversion[11,12,13,14,15,16,17]. Yet, CT kinetics are strongly influenced by large-scale fluctuations that modulate inter-fragment geometry and electronic delocalization. These effects renormalize the balance between electronic coupling and the vibrational landscape, thereby blurring the boundary between adiabatic and non-adiabatic regimes[8,18].

While the original Marcus theory captures thermally averaged CT activation, its vibronic extension within the Marcus–Levich–Jortner (MLJ) framework predicts a structured dependence of the CT rate on driving force, including discrete high-frequency modes and resonant channels in both normal and inverted regimes[19]. So far, CT rate modulations have never been experimentally evidenced mostly due to environmental averaging that usually prevents a clear picture of intrinsic vibrational states[20,21,22,23].

At its most fundamental level, CT is intrinsically governed by electronic coupling and nuclear reorganization, key parameters that are primarily defined at the molecular scale[24]. Gaining access to these parameters is therefore essential to decipher the interplay between adiabatic and non-adiabatic regimes. However, obtaining reliable information about CT at the molecular scale remains highly challenging, as the relevant variables are rarely controlled independently, masking the underlying microscopic pathways. Resolving CT at the nanoscale under well-controlled conditions is thus a prerequisite for establishing a predictive description of CT mechanisms.

Although CT has been extensively studied in solution and bulk systems[11,25,26], intermolecular CT between individual molecules immobilized on solid surfaces remains rarely explored[27,28,29]. Ensemble measurements on surface-confined assemblies[30,31] often conform to original Marcus-type behavior, while averaging over molecular orientation, local structure, and specific pathways, concealing single-pair dynamics[32,33,34].

Progress has been hindered by the substrate footprint: metallic surfaces induce strong hybridization and ultrafast relaxation, whereas bulk insulating surfaces may introduce trapping and morphological disorder[35,36,37]. The absence of a weakly coupled, structurally defined platform has thus prevented quantitative benchmarking of intermolecular CT at the single-dyad level.

Here, we quantify the respective roles of adiabatic and non-adiabatic CT contributions by suppressing environmental fluctuations, eliminating ensemble averaging, and enabling independent control over key parameters using a molecular device composed of two $H_2$TPP molecules assembled in situ physisorbed on $CaF_2$/Si(100) (Fig. 1a)[38,39]. The dyad fixes intermolecular separation while suppressing diffusion. CT is locally initiated by STM hole injection, with no measurable substrate-mediated contribution[40]. Charge recombination induces a binary tautomerization signature detectable by STM[41,42]. Performed at 9 K, this approach enables direct and quantitative access to CT statistics governed by quantum and vibronic dynamics.

**Adsorption conformations and electronic structure of the $H_2$TPP and ($H_2$TPP)$_2$ dyad.**

Following adsorption of $H_2$TPP on the $CaF_2$/Si(100) surface, STM imaging reveals several molecular conformations distributed along the $CaF_2$ stripe ribbons. Among them, two dominant adsorption geometries were identified (Figs. 1b-e). Both configurations are positioned between adjacent $CaF_2$ stripes and display distinct central features arising from the porphyrin core.

Figure 1b shows the first configuration, in which the two inner hydrogen atoms adopt an up-down arrangement (denoted $H_2$TPP-UD). In contrast, the second conformation (Fig. 1c) differs only by the orientation of the inner hydrogens, which are both directed upwards ($H_2$TPP-AU).

Density functional theory (DFT) calculations of the occupied density of states (DOS) (Figs. 1d,e) reproduce the experimental contrast and support the structural assignment of both configurations. A detailed structural analysis, including height profiles, is provided in Supplementary Fig. S1 and Supplementary Note 1.

Note that these conformations in which the $H_2$TPP molecule loses $D_2$h symmetry have been suggested but have never yet been clearly identified[43,44]. The positions of the two inner hydrogen atoms are confirmed by telegraphic-noise measurements across the four pyrrole units, with elevated tunneling current noise specifically observed at the hydrogen-bearing groups (Supplementary Fig. S2). DFT calculations show that $H_2$TPP-AU is slightly more stable than $H_2$TPP-UD (~0.05 eV) and more stable than the planar gas-phase geometry[45]. The $H_2$TPP-AU conformation was therefore chosen for the CT investigation.

The electronic structure of the $H_2$TPP-AU molecule is described in Fig. 2a via dI/dV spectroscopy curves that probe the occupied and unoccupied states of the physisorbed molecules from -2.7 V to +2.5 V. These curves exhibit a sharp dI/dV peak lying between -1.95 V and -2.05 V and a larger peak centered at -2.6 eV. Both peaks are observed at various selected measurement locations (points 1 to 9 in the inset in Fig. 2a) and they distinctly reveal a delocalization of the dI/dV peaks over the entire molecule. Note that due to the use of the $CaF_2$ insulating layer, one can observe that almost no unoccupied states are measured in the dI/dV curves between -1.8 V and 2.5 V, illustrating the strong electronic decoupling between the molecule and the $CaF_2$/Si(100) substrate[38].

The delocalized dI/dV peaks across the entire molecule are explained in Fig. 2b, which compares the gas-phase (no surface) DFT calculation of the $H_2$TPP-AU conformation with the calculated projected density of states (PDOS) of the $H_2$TPP-AU molecule adsorbed on the $CaF_2$/Si(100) surface (right panel in Fig. 2b). The electronic structure of the $H_2$TPP-AU molecule includes a small set of frontier orbitals in which the DOS is primarily distributed across the porphyrin macrocycle (occupied and unoccupied pyrrole groups in Fig. 2b).

A secondary set of orbitals contributes to the occupied and unoccupied dI/dV peaks associated with DOS localized on the pyrrole and phenyl groups. As shown by the PDOS curves (Fig. 2b, right panel), both groups exhibit similar trends with two sets of separated peaks. This DOS delocalization is primarily governed by the dihedral angle between the phenyl rings and the porphyrin macrocycle, which tunes the electronic coupling within the molecule core[40]. Under these conditions, the calculated HOMO–LUMO gap of $H_2$TPP–AU is ~3.9 eV in the gas phase and ~1.5 eV upon adsorption on $CaF_2$/Si(100).

Upon formation of the ($H_2$TPP-AU)$_2$ dyad (Fig. 2c and Supplementary Fig. S3), dI/dV spectroscopy was performed at three sites ($P_1$–$P_3$) on the lower molecule (Fig. 2c, right panel). These sites were selected to trigger meso-N tautomerization in the upper molecule via CT. The resulting dI/dV signals, which are nearly identical across all probed positions (Fig. 2d), were recorded at decreasing tip–surface distances $dz$, where $dz = z_0 - \delta z$. Note that the reference measurements in Fig. 2a were taken at $\delta z = 0$, $z_0$ being defined by the initial tunneling setpoints.

Reducing the initial tip height value by $\delta z = 1$ Å (top panel, Fig. 2d) reveals a new dI/dV peak at −1.78 V. This feature corresponds to a positive ion resonance ($PIR_1$) and is accompanied by vibrational fine structure between the two dI/dV peaks at −2.1 V and −1.78 V, indicating the onset of vibronic modes of the excited $H_2$TPP cation[46]. Further decreasing the tip–surface distance to $\delta z = 1.5$ Å (central panel, Fig. 2d) yields an additional ion-resonance peak at −1.87 V ($PIR_2$). At $\delta z = 1.8$ Å, these features are confirmed, while the initial peak at −2.1 V is significantly reduced. The observed PIRs correspond to transient molecular electronic states that appear positively charged due to tunneling through the double-barrier junction between the surface and the STM tip[47], without the formation of a stationary molecular ionic species, consistent with previous observations on this surface[38,40].

The dI/dV spectra acquired between the $PIR_{1/2}$ energies at a constant δz = 1.2 Å for the excitation point $P_1$ were precisely examined in Fig. 2e over the bias range −1.7 to −2.5 V. The $PIR_1$ peak energy is used as the reference for assigning vibrational energies of the probed $H_2$TPP-AU cation (Supplementary Fig. S4 for additional examples). The observed modes fall into two main groups. The first low-frequency range spans ~47−737 $cm^{-1}$ and is primarily associated with phenyl torsional motions and macrocycle deformations. The intermediate region spans ~832−1613 $cm^{-1}$ and comprises collective molecular stretching modes of the porphyrin framework, including contributions from the central N−H units coupled to pyrrole and phenyl breathing motions. This includes possible in-phase and out-of-phase contributions of the two central hydrogen atoms within the dyad**.** As the surface bias approaches the second PIR voltage at −1.87 V, additional high-wavenumber modes emerge between 1814 $cm^{-1}$ and 1927 $cm^{-1}$, likely consistent with overtone or combination modes activated by vibronic coupling. At even higher frequencies (~ 3200 $cm^{-1}$), one expects N−H and C−H stretching modes, although these were not measured.

The origin of these vibrational modes was elucidated by calculating the vibrational spectra of an isolated $H_2$TPP-AU and $(H_2TPP\text{-}AU)_2$ dyad in their neutral and cationic states (Supplementary Fig. S5). While the dyad naturally possesses a larger vibrational manifold, the calculations reveal the emergence of additional modes that are absent in the isolated molecule and arise from intermolecular coupling. Upon cation formation, the dyad exhibits enhanced vibrational intensity in the 1400−1700 $cm^{-1}$ range compared with the 500−1000 $cm^{-1}$ in the neutral state, indicating a redistribution of vibronic activity toward higher-energy modes. All measured frequencies (Supplementary Fig. S4) are in reasonable agreement with calculated values (Supplementary Fig. S6). Analysis of the most intense modes on each pyrrole reveals nearly identical counts (15 at $P_1$ and $P_3$; 14 at $P_2$), with modes predominantly involving collective stretching and torsional contributions of the porphyrin framework, including the inner N−H units, indicating delocalized vibrational populations across the non-covalent dyad (Supplementary Figs. S7–S9).

**CT measurements in the $H_2$TPP dyad.**

To study electronically induced CT in the $(H_2TPP)_2$ dyad, we selectively excite three previously identified sites on the pyrrole groups of the lower (L) $H_2$TPP-$AU_\alpha^L$ molecule (Fig. 3a). The excitation positions were chosen to prevent molecular displacement on the surface and promote local cation formation, as described in previous work[40,48]. Voltage pulses of varying amplitudes are applied at random to the three selected positions of the $H_2$TPP-$AU_\alpha^L$ molecule of the dyad to trigger the tautomerization process in the $H_2$TPP-$AU_\alpha^U$. After each voltage pulse, the central region of the upper $H_2$TPP-$AU_\alpha^U$ molecule in the dyad is imaged with the STM to detect any structural changes within the dyad (Fig. 3a). When tautomerization occurs, the upper molecule conformation changes to $H_2$TPP-$AU_\beta^U$ (Fig. 3b) and serves as a clear signature of charge transfer within the dyad (see method and Supplementary Fig. S10). Our DFT calculations indicate that the observed conformation is relative to a meso-N tautomerization which energy lies between the $H_2$TPP-$AU_\alpha$ and the cationic $H_2$TPP-$AU_\alpha$ state (Supplementary Figs. S11, S12 and Note 2).

The CT process in the dyad is investigated as a function of the tunneling electron energy by repeating multiple excitation measurements across a bias range of −2.0 V to −2.9 V in 0.1 V increments (Fig. 3c). The CT rate, extracted from the probability of molecular change due to the meso-N tautomerization is plotted as a function of the excitation bias for each considered point (see Methods). The evolution of the resulting CT rates is presented in Fig. 3c (blue circles). The oscillatory dependence of the CT rate on tunneling-electron energy shows a dominant maximum at −2.3 V, common to all excitation sites, consistent with a delocalized molecular resonance or enhanced electronic coupling. Additional site-specific extrema, at higher at $P_1$ and lower energies, at $P_2$ and $P_3$, point to secondary mechanisms that merit further investigation.

To understand the evolution of the CT rates presented in Fig. 3c, we first examined the influence of the electronic coupling between the two fragments of the $(H_2TPP\text{-}AU)_2$ dyad. This analysis utilized a Fermi-Golden rule-like expression[40]. The CT rate can thus be expressed as:

$$\Gamma(E) = \frac{2\pi}{\hbar} \underbrace{\sum\nolimits_{i \in initial} P_i(\varepsilon_i) f_I(\varepsilon_i - \mu_I)\delta(E - \varepsilon_i)}_{Excited} \underbrace{\sum\nolimits_{j \in final} \theta(\varepsilon_j - \varepsilon_i) f_F(\varepsilon_j - \mu_F) \left|V_{ij}\right|^2}_{Neutral} \quad (1)$$

Where E is the tunneling energy of the electron leaving the excited molecule, $P_{i=1,2,3}(\epsilon_i)$ is the local partial density of state PDOS at the points $P_1 - P_3$, respectively. The terms, $f_I$ and, $f_F$ are the Fermi-Dirac functions pertaining to the initial and final states, $\theta(\varepsilon)$ is the Heaviside step function at the energy $\varepsilon$ that ensures thermodynamic irreversibility. Here, $\left|V_{ij}\right|^2$ stands for the electronic coupling connecting the electronic states of donor and acceptor. The chemical potential of each fragment $\mu_{I/F}$ is related to the Fermi Energy, $E_F$, $\varepsilon_{i/j}$ are the values of the energy levels and $\delta(\varepsilon)$ is the Dirac delta function at the energy $\varepsilon$ (Supplementary Note 3 and the Methods section).

The variations of Γ(E) computed from Eq. (1) are shown in Fig. 3d,e for two cases. Figure 3d reports $\Gamma_{neutr.}(E)$ for both $H_2TPP\text{-}AU_\alpha$ fragments in their neutral states, over an energy range of 0–4 eV below the surface Fermi energy level. For $P_1$–$P_3$, $\Gamma_{neutr.}(E)$ exhibits a principal maximum near 2.3 eV, with peak amplitudes ranked $P_1$ ($2.5 \times 10^{10}$ $s^{-1}$) > $P_2$ ($6.5 \times 10^{9}$ $s^{-1}$) > $P_3$ ($4.3 \times 10^{9}$ $s^{-1}$), consistent with the experimental ordering at −2.3 V (Fig. 3c).

However, this approach does not capture the experimentally observed periodic modulation of the CT rate. We therefore computed $\Gamma_{cat.}(E)$ for cationic $H_2$TPP-AU$_\alpha^L$ over −2 to −3.5 eV below the Fermi level (Fig. 3e), revealing a maximum at −2.3 eV (1.5 × $10^{10}$ $s^{-1}$) and weaker local maxima at −2.6 eV (2.3 × $10^9$ $s^{-1}$) and −2.8 eV (6 × $10^8$ $s^{-1}$). Although some features align with experiment, the relative amplitudes and absence of periodicity indicate additional processes beyond the cationic Fermi–Golden-Rule description of Eq. (1).

Guided by the observation that positive-ion resonances drive vibrational activation, we use the MLJ model to include vibronic effects in the CT process[48]. Accordingly, the CT-rate expression is:

$$\Gamma_{MLJ}(E) = \frac{2\pi}{\hbar}\left|V_{ij}\right|^2 \frac{1}{\sqrt{4\pi\lambda_s k_B T}} \sum_{n=1}^{\infty} e^{-S_e} \frac{S_e^n}{n!} exp\left\{-\frac{(\Delta G_T+\lambda_s+n\hbar\omega_e)^2}{4\lambda_s k_B T}\right\} \qquad (2)$$

Here, $\lambda_s$ denotes the dyad's reorganization energy, accounting for the surface environment in the CT process. The electronic coupling between the two $H_2$TPP-AU$_\alpha$ molecules, $\left|V_{ij}\right|^2$, is analogous to that obtained in Equation (1). $\Delta G_T$ is the change in the total Gibbs free energy between initial and final states, and $\hbar\omega_e$ is the frequency of an effective vibrational mode in the dyad. The Huang–Rhys factor, $S_e = \lambda_i/\hbar_e\omega_e$, quantifies the overlap between the reaction coordinate and vibrational mode $n$ in the CT rate, where $\lambda_i$ is the dyad's internal reorganization energy.

Under our STM conditions, the total Gibbs free-energy change is set by the bias-driven force $qV_{exc.}$ ($q$ being the electron charge). Sweeping $V_{exc.}$ = -1.7 V to $V_{exc.}$ = -2.9 V during CT-rate measurements extracts an electron from excited $H_2$TPP-AU$_\alpha^L$ and yields $\Delta G_T = \Delta G_0 + \gamma q V_{exc.}$ (Fig. 3f), where $\Delta G_0$ is the energy to reach the local adiabatic ionization potential (IP), and γ the fraction of the driving force applied beyond $\Delta G_0$. Tuning the tip–sample bias thus locally controls cation formation, triggers CT, and activates specific vibrational modes.

Using equation (2), we fit the experimental data with a sum of Gaussians (red curves in Fig. 3c) to capture key CT features in the $(H_2TPP\text{-}AU)_2$ dyad. The procedure is detailed in Supplementary Fig. S13 and Notes 4. The fitted parameters are listed in Table 1. In this framework, the inter-peak spacing in Fig. 3c equals the vibrational energy $\hbar\omega_e$ (n = 0 − 1 at $P_1$ and $P_3$, n = 0 − 2 at $P_2$), while each peak full width at half maximum (FWHM) scales as $\sqrt{2\lambda_s k_B T}$, set by the reorganization energy $\lambda_s$ and the system temperature.

The excitation energies $\Delta G_{Max}$ at which the CT rate peaks are maximum at $\Gamma_{max}$, the corresponding FWHM values, and the vibrational energy $\hbar\omega_e$ given by the differences between two successive $\Delta G_{Max}$ values, can be directly extracted from Fig. 3c (Table 1). The vibrational energies, derived from the energy difference between two consecutive CT maxima, are 1676 $cm^{-1}$ at $P_1$, 1613 $cm^{-1}$ and 2016 $cm^{-1}$ at $P_2$, 2339 $cm^{-1}$ at $P_3$. Comparison with vibrational spectra confirms that these modes are consistent with the experimentally and calculated vibrational modes (Supplementary Figs. S4, S5 and S6).

Numerical fitting of the data with Gaussian series functions (equation 2) allows extraction of additional parameters (Table 1, lower part). Fixing $\Delta G_0$ to the PIR energy from Fig. 2d (−1.78 eV) yields the reorganization energy $\lambda_s$, the Huang–Rhys factors $S_e$, the internal reorganization energy $\lambda_i$ and the electronic coupling $|V_{ij}|^2$.

The reorganization energy $\lambda_s$ increases from 0.32 eV at $P_1$ to 0.54 eV at $P_3$, indicating its correlation with the locally induced charge-transfer process. In contrast, the Huang–Rhys factor $S_e$ is markedly larger at $P_1$ than at $P_2$ and $P_3$, reflecting stronger vibrational coupling at $P_1$. The higher internal reorganization energy $\lambda_i$ at $P_1$ suggests that the CT dynamics are optimized for $\hbar\omega_e$ = 1676 $cm^{-1}$, whereas at $P_2$ and $P_3$ the process involves a broader set of vibrational modes, consistent with the emergence of higher-energy CT peaks at $P_2$. The effective thermal broadening (63–290 K) scales with linewidth and exceeds the contribution expected from the single resonant mode $\hbar\omega_e$, suggesting additional unresolved vibrational channels in the CT process.

Because $|V_{ij}|^2$ is fitted for each Gaussian component, it becomes energy dependent and increases modestly with $n$. Table 1 shows relatively small $|V_{ij}|^2$ at $P_1$, with higher values at $P_2$ and $P_3$ that track increases in $\lambda_s$. Linear regressions of $|V_{ij}|^2$ versus energy yield slopes of 24, 96, and 145 $cm^{-1}.eV^{-1}$ for $P_1$–$P_3$, respectively, indicating that higher-energy CT channels involve stronger effective coupling and enhanced environmental reorganization.

A careful analysis of our fitting data allows several additional robust trends to be identified. A central observation is that the fitted electronic prefactor $|V_{ij}|^2$ increases from $P_1$ to $P_3$ while the CT efficiency remains maximal at $P_1$. Within MLJ theory, the combined balance of electronic coupling, Franck–Condon (FC) statistics, and energetic resonance governs the rate, so a larger $|V_{ij}|^2$ does not necessarily yield enhanced CT rate. These results demonstrate that charge-transfer efficiency is not governed by electronic coupling alone but emerges from a balance between electronic and vibronic contributions.

Instead, CT rate at $P_1$ is best interpreted as being strongly supported by nuclear reorganization, as the Huang–Rhys factor $S_e$>1 produces a relatively broad FC distribution capable of compensating weaker electronic coupling. As the internal reorganization $\lambda_i$ remains rather within a relatively narrow range (i.e. $\langle\lambda_i\rangle$ ~ 0.27 eV) across the excitation positions, the dominant CT evolution arises from an increase of the effective reorganization energy contribution $\lambda_s$, which follows the growth of the effective vibrational pseudo-frequency $\hbar\omega_e$. Together, these trends suggest that the CT reaction coordinate becomes increasingly shaped by environmental degrees of freedom (substrate) while the initial intramolecular nuclear framework is not the sole contribution to the evolving reaction coordinate. Therefore, we show that spatially varying the excitation location ($P_1$ to $P_3$) primarily re-weights the relative vibronic and electronic contributions governing the CT rate within the MLJ description.

Within the Marcus–Levich–Jortner description, the concurrent observation of $S_e$<1 with larger fitted vibrational quanta, and increased $|V_{ij}|^2$ at specific excitation positions reflects a redistribution of how nuclear and electronic factors contribute to the MLJ rate at each excitation point. Because $S_e=\lambda_i/(\hbar\omega_e)$, an increase in the effective vibrational quantum reduces the Huang–Rhys factor while preserving a comparable internal reorganization energy $\lambda_i$. As a result, the FC envelope therefore narrows, reflecting a reduction of the mean vibronic order $\langle n \rangle = S_e$ at nearly constant $\lambda_i$. It is hence crucial to understand that the vibronic manifold does not correspond to a smaller nuclear reorganization. Rather, the same internal reorganization energy is distributed over fewer, higher-energy quanta.

Taken together, these observations indicate that CT proceeds from a non-thermalized vibronically excited cationic configuration generated upon local electron injection, rather than from the fully relaxed frontier SOMO orbital. In this sense, the process can be viewed as an electronic analogue of anti-Kasha behavior[49], in which the reactive pathway originates from deeper hole states (HOMO−n) rendered accessible through vibronic coupling. In our case, this scenario suggests Herzberg–Teller–type activation arising from symmetry breaking in the non-planar $(H_2TPP–Au)_2$ dyad (Figs. 1c, S1, S11), which enables vibronic coupling between otherwise weakly allowed electronic configurations. These observations thus reflect a spatial redistribution of nuclear and electronic contributions within a common CT mechanism, indicating the presence of position-dependent CT channels whose nanoscale origin remains to be clarified.

**Discussion**

Interpreting CT processes on surfaces requires identifying the key governing parameters. Two reaction coordinates are considered: $d$, the initial intermolecular distance within the dyad, following the Marcus framework, and $h$, the initial dyad–substrate interaction height, following the Antoniewicz model. Specifically, $d$ denotes the center-to-center distance between the two $H_2TPP-AU_\alpha$ molecules, while $h$ corresponds to the molecular height above the surface (Fig. 4a). Their evolution during CT is shown in Figs. 4a–d. This framework merges 2D features of the Antoniewicz model described by a Morse potential (Figs. 4e–g) and the Marcus model, using harmonic potentials, (Figs. 4h–j), with parameters evolving along the $h$ and $d$ coordinates, respectively.

Before excitation (Fig. 4a), the two molecules of the $(H_2TPP-AU_\alpha)_2$ dyad, $M_1$ and $M_2$, are physisorbed at a center-to-center distance $d_0$ and at the same height $h_0$ above the surface. Upon electronic excitation of $M_1$, forming a local cation ($M_1^+$), the molecule shifts toward the surface (Fig. 4b). This vertical displacement arises from image-charge attraction, reducing the molecule–surface distance from $h_0$ to $h_c$[50,51]. Geometrically, it also increases the intermolecular separation from $d_0$ to $d_c$.

The introduction of an image charge in the first excitation step is consistent with cation formation on the $CaF_2$/Si(100) surface, as described by the Antoniewicz process (Fig. 4e). This shifts the Morse-like potential minimum of the dyad in its cationic state along the $\vec{h}$ axis (blue and yellow curves, Fig. 4e). The reorganization energy $\lambda_c$, defined as the energy difference between the cationic Morse potential at $h = h_0$ and its minimum, quantifies this process. From the ionization point ($h = h_0$), the dyad evolves along the cationic Morse potential toward its minimum at $h = h_c$, corresponding to a vertical displacement $\Delta h_c = h_0 - h_c$. A related Marcus diagram (Fig. 4h) can also be constructed to include the lateral shift from $d_0$ to $d_c$, where the harmonic potential of the $(H_2TPP\text{-}AU^+)_2$ dyad is displaced by $\Delta d_c = d_c - d_0$ (blue and yellow curves, Fig. 4h). The relationships between $\Delta h_c$ and $\Delta d_c$ are detailed in Supplementary Notes 5 and 6.

In Fig. 4h, $\Delta \mathrm{d_c}$ denotes the maximum displacement along the d axis at the largest vertical shift $\Delta h_c$ while the nascent dynamics of one fragment of the $(H_2TPP\text{−}AU^+)_2$ dyad evolves immediately following the cation formation near the vertical transition at $h = h_0$ (Fig. 4e). The 2D curves in Fig. 4h therefore represent harmonic potentials at two different $h$ values along the Morse potential, with $\Delta d_c$ spanning from 0 to $d_c$ while $\Delta h_c$ varies. The Gibbs free energy $\Delta G_0$ between these harmonic potentials is thus defined at the instant of the cation formation ($h = h_0$ and $\Delta d_c = 0$ in Fig. 4h). Accordingly, the reorganization energy $\lambda_c$ is not directly associated with the $d$ coordinate but instead constitutes part of the external reorganization energy imposed by the surface.

The next step in the CT process occurs when the neutral $H_2TPP\text{-}AU_\alpha^U$ ($M_2$) fragment transfers an electron to neutralize its cationic $H_2TPP\text{-}AU^+{}_\alpha{}^L$ counterpart ($M_1^+$), producing a stable charge inversion (Fig. 4c). In this step, the height $h_T$ of the newly positively charged fragment may differ slightly from $h_c$, as the CT event perturbs the entire charge density of the neutral $H_2TPP\text{−}AU_\alpha^U$. This induces a shift $\Delta h_T$ in the Morse potential minima of the two cationic states, and a corresponding change $\Delta d_{CT}$ in the minima of the associated Marcus harmonic curves (Figs. 4f, 4i). Because the two dyad molecules are structurally identical, the Gibbs free-energy difference between the cationic configurations can be considered negligible, yielding two degenerate harmonic potentials (Fig. 4i). Under these conditions, the reorganization energy $\lambda_s$ used in equation (2) is defined between the two cationic states through their harmonic potentials (Fig. 4i). The relation between $\Delta h_T = h_T - h_c$ and $\Delta h_t = h_0 - h_T$ is described in Supplementary Notes 5.

The final step of the CT process (Fig. 4d) involves a conformational change of $H_2$TPP-$AU_{\alpha}^{U+}$ in its cationic state through a meso-N tautomerization, yielding $H_2$TPP-$AU_{\beta}^{U+}$. The resulting ($H_2$TPP−$AU^+$)$_2$ dyad is then neutralized, stabilizing the system (Figs. 4g, j). These events substantially lower the dyad's Gibbs free energy, as the system nearly returns to its initial configuration through the meso-N tautomerization of $H_2$TPP-$AU_{\beta}^{U}$ (Supplementary Fig. S11, S12).

STM-tip excitation localizes the initial cation at a well-defined site on a pyrrole group of the lower $H_2$TPP-$AU_{\alpha}^{L}$ molecule, consistent with the HOMO/SOMO distribution that appears confined at the excited pyrrole units (insets in Figs. 3d, e). By contrast, the neutralizing electron density from $H_2$TPP-$AU_{\alpha}^{U}$ molecule may be less localized, allowing the image-charge potential minima at $\mathrm{h_T}$ to differ from $\mathrm{h_c}$. To capture this effect, we model the area from which the electron is transferred as a delocalized charge density volume over a sphere of radius R. The ensuing Morse-potential minimum $h_T$ therefore differs from $h_c$ so that the shift $\Delta\mathrm{h_T}$ is a function of $h_0$ and R (Supplementary Note 5, 6).

To emphasize this mechanism, we analyzed the energy alignment and spatial overlap between orbitals localized at the excited pyrrole sites ($P_1$−$P_3$) of $H_2$TPP-$AU^{+}{}_{\alpha}^{L}$ and those distributed across the $H_2$TPP-$AU_{\alpha}^{U}$ fragment. Spin-polarized Kohn−Sham (KS) energies and spatial distributions of the first 55 occupied orbitals in the cationic dyad (0−5.5 eV) were compared (Supplementary Figs. S14 and Note 7). The initial spin state of the local cation, namely whether the SOMO-n is spin-up or spin-down, in $H_2$TPP-$AU^{+}{}_{\alpha}^{L}$ defines a distinct orbital landscape that influences energetic matching between fragments. A few spin-up orbitals at $P_2$ and $P_3$ align with those of the upper fragment ($H_2$TPP-$AU_{\alpha}^{U}$), whereas at $P_1$, several spin-down states coincide energetically. The high density of orbitals within a narrow energy window (−2.8 to −2.1 eV) at the upper-fragment indicates the presence of multiple channels spatially delocalized for charge inversion upon cation formation at $H_2$TPP-$AU^{+}{}_{\alpha}^{L}$, independently of the spin state of the upper $H_2TPP\text{-}AU_\alpha^U$ fragment. This behavior is consistent with a delocalized charge-inversion process as described earlier.

Our investigation of CT processes in an $(H_2TPP)_2$ dyad adsorbed on a semi-insulating surface provides a unified framework that integrates the Marcus (MLJ) and Antoniewicz models to capture complementary aspects of hole-transfer dynamics. To fully clarify the coupling between these models, we construct a three-dimensional potential energy surface (PES) defined by the key parameters $h$ and $d$ (Fig. 5). Sectional planes perpendicular to the $\vec{d}$ or $\vec{h}$ axes yield 2D Morse and harmonic potentials, respectively (Fig. 5a).

Vertical ionization drives the $(H_2TPP\text{-}AU_\alpha)_2$ dyad into a local cationic state, corresponding to a transition from the blue to the orange 3D PES (Fig. 5a). At this initial stage, neither d nor h has changed, and both Marcus harmonic potentials share the same minimum at $h = h_0$. The system then evolves quantum dynamically as the cationic fragment moves toward the surface, producing shifts $\Delta h$ and $\Delta d$ that may ultimately reach their maxima, $\Delta h_c$ and $\Delta d_c$, respectively (Fig. 5b). These coordinate changes generate two distinct potential profiles in the 3D PES (blue and orange dotted curves, Fig. 5b), consistent with the 2D representation in Fig. 4h.

As the excitation bias further decreases to Vs << 0, additional energy ($\gamma q V_{exc.}$) is imparted to the molecular system, driving the ionized molecule into a distribution of vibronic modes across different energies (Figs. 3f, 5c). Here, the FC overlap determines which vibrational modes in the first cationic state $(H_2TPP\text{-}AU_\alpha^L)^+$- $(H_2TPP\text{-}AU_\alpha^U)$ Morse potential curve can couple to the second $(H_2TPP\text{-}AU_\alpha^L)$ - $(H_2TPP\text{-}AU_\alpha^U)^+$ . The vibrational population prepared at ionization can therefore promote charge inversion through FC-allowed transitions. Simultaneously, the onset and rate of inversion are influenced by the local slope and curvature of the Morse potential near $h_0$, which direct nuclear motion toward the cationic minimum. Together, these factors FC weighted coupling and PES driven forces, govern the early CT dynamics, with the displacements $\Delta h$ and $\Delta d$ emerging as key observables. These results indicate that charge transfer on surfaces cannot be captured by a single reaction coordinate but instead requires a multidimensional description coupling intermolecular and surface degrees of freedom.

Our model, which integrates a 3D Morse–Marcus potential energy surface (PES) along the $h$ and $d$ coordinates, allows the reorganization energy $\lambda_s$ (Fig. 4i) to be expressed directly in terms of the system parameters. A key geometrical feature of the Marcus harmonic potential is that $\lambda_s$ scales with the displacement $\Delta d_{CT}$ where $\Delta d_{CT} = k\sqrt{\lambda_s}$, where $k$ is the harmonic potential constant, assumed identical across all 3D PESs (Supplementary Notes 5, 6). Thus, $\lambda_s$ can be written as:

$$\lambda_s = \frac{1}{k}\left(\sqrt{d_0^2 + \left(A\frac{1}{2\kappa^2\lambda_i}\frac{R^2}{5h_0^4}\right)^2} - \sqrt{d_0^2 + \left(A\frac{1}{2\kappa^2\lambda_i}\frac{1}{h_0^2}\right)^2}\right)^2 \quad (3)$$

Here, κ is a dimensional factor related to the width of the Morse potential, and A = $Q^2/(16\pi\epsilon_0)$. Equation (3) shows that the reorganization energy $\lambda_s$ associated with charge inversion, as described by the MLJ equation (2), depends on the initial molecular height $h_0$, the initial intermolecular separation $d_0$, and the internal reorganization energy $\lambda_i$, introduced previously through the Huang–Rhys factor $S_i = \lambda_i/\hbar\omega_e$. R is the radius of the charge-density sphere defined earlier. Using equation (3), the measured trends in Table 1 were reproduced by fitting $\lambda_s$ as a function of $h_0$ (Fig. 5d). The ball-and-stick molecular inset in Fig. 5d reports the DFT-derived $h_0$ values at the three excitation sites ($P_1$–$P_3$), plotted along the abscissa of the $\lambda_s(h_0, d_0, R, \lambda_i)$ graph.

The red fitted curve in Fig. 5d reproduces the experimental trend with $\lambda_i$ = 0.3, R = 5.6 Å, κ = 1 Å$^{-1}$, and A = 3.6 eV·Å, giving $\lambda_s$ = 0.34, 0.49, and 0.51 eV at $P_1$–$P_3$ for which $h_0$ = 6, 5, and 4 Å, respectively (see the inset in Fig. 5d). Here, $d_0$ (fixed at 14.6 Å, Supplementary Fig. S3) sets the $\lambda_s$ maximum. The fitted radius R matches the tetraphenyl-porphyrin size (Fig. S1a) and further indicates that a large DOS fraction of the $H_2TPP\text{-}AU_\alpha^U$ fragment contributes to charge inversion. Note that for $h_0 < R/\sqrt{5}$ (~2.5 Å), $\lambda_s$ rises sharply due to enhanced image-charge effects, corresponding to a chemisorption regime not reached experimentally.

Building on the MLJ analysis above, the combined Antoniewicz–Marcus model provides a dynamical interpretation of the observed trends, highlighting the dual role of the initial excitation site and image-charge–driven nuclear motion. As shown in Table 1, the relatively small values of the electronic coupling parameter $V_{ij}$ alone cannot account for the observed CT rates within a purely electronic framework. Instead, weaker electronic coupling correlates with an increased Huang–Rhys factor $S_e$, shifting the CT contribution toward vibronically assisted pathways.

Within a Morse description of the cationic potential, the width parameter $a$ determines the local curvature of the potential energy surface and thus sets the nuclear timescale along the CT coordinate (Supplementary Note 5). A smaller $a$ corresponds to a shallower curvature and therefore to slower nuclear motion, extending the temporal window during which the excited cation evolves on its potential energy surface. This modified dynamical regime alters the timescale over which non-adiabatic coupling can act, thereby influencing the contribution of vibronically assisted CT channels as $\gamma q V_{exc.}$ increases. Near the potential minimum, a harmonic expansion gives $a \propto \sqrt{\lambda_i / D_e}$. Assuming that $D_e$ (i.e the dissociation energy in the Morse potential) remains constant upon cation formation, this scaling indicates that the local curvature, and therefore the nuclear timescale is influenced by the internal reorganization energy $\lambda_i$.

Equation (2) also indicates that the reorganization energy $\lambda_s$ must remain relatively small to achieve high CT rates in the modulation regime of Γ(E) (Supplementary Fig. S15 and Note 8). The graphical representation of Equation (3) in Fig. 5d shows that optimizing Γ(E) via $\lambda_s$, can be accomplished in two ways: either by increasing $h_0$ toward larger values, corresponding to dyad configurations approaching the gas phase, or by tuning $h_0$ close to $R/\sqrt{5}$. The latter case highlights the role of the separation between the two cation Morse potentials, quantified by the shift $\Delta h_T$ (Fig. 4f), which defines the corresponding distance $\Delta d_{CT}$ (Fig. 4i). Hence, optimizing $\Gamma(E)$ also implies tuning with these parameters while it also requires aligning the vibronic population of the dyad to promote meso-N tautomerization via charge inversion.

Assuming $\Delta G_{CT}$ remains negligible compared to $\lambda_s$, an increase in $\lambda_s$ translates into a larger separation $\Delta d_{CT}$ between the potential energy surfaces (Fig. 4i). While this shift moves the overlap of harmonic potentials to higher energies, which would classically reduce the CT rate, the enhanced coupling to high-frequency modes facilitates charge inversion through vibronic activation. This further opens anti-Kasha-like CT pathways, a mechanism supported by our experimental data where the resonant excitation frequencies align with the high-energy vibronic spectrum of the $(H_2TPP\text{-}AU)_2$ cation. This correlation confirms that the charge inversion is driven by these energetic vibrational states, which trigger a charge-induced tautomerization analogous to excited-state intramolecular proton transfer[52]. In this framework, the proton shift acts as the structural relaxation response to the new electronic landscape, providing a quantitative signature of the charge transfer event.

This approach establishes a direct link between the experimentally extracted reorganization energy and the molecular geometry, making $\lambda_s$ a predictive quantity rather than a fitted parameter. Within this framework, charge transfer at surfaces emerges from a multidimensional energy landscape in which electronic coupling, vibronic activation, and surface-induced reorganization are intrinsically coupled. This description reveals two distinct regimes for optimizing charge transfer: either by increasing the molecule–surface separation leading to conditions similar to diluted media, or by tuning the system toward an image-charge dominated regime analogous to solvation.

Optimizing Γ(E) therefore requires controlling these geometric parameters, which directly govern the balance between electronic and vibronic contributions to the CT process. This framework bridges surface science and gas-phase descriptions and is readily extendable to other excitation pathways, most notably photonic excitation. In such a scenario, the initial cationic state would be replaced by an optically excited state associated with a well-defined electronic transition, following a strategy reminiscent of the Menzel-Gomer-Redhead (MGR) model[53]. Such an extension naturally encompasses hot-electron generation within the substrate and its subsequent coupling to adsorbed molecules, opening a compelling direction for future studies.

## METHODS

### Surface and adsorbate preparation

The experiments are performed with a low-temperature (9 K) scanning tunneling microscope (STM) working in ultra-high vacuum (UHV) from Createc using a PANSCAN scanner head[54]. The surfaces are prepared from highly doped (n-type, As-doped, $\rho$ = 5 mΩ.cm) Si(100) samples from ITME. The bare silicon surface is reconstructed in a c(4×2) arrangement as explained in several previous studies via multiple annealing cycles[55]. To minimize surface defects, the base pressure in the UHV is maintained below $4\times10^{-11}$ Torr during this process and the maximum temperature during rapid heating does not exceed 1120 °C, preventing excessive dopant diffusion toward the silicon surface. A thin layer of $Ca_xF_y$ is then deposited by exposing the bare silicon surface to sublimating $CaF_2$ molecules, while maintaining the silicon surface at approximately 750°C. The evaporation of the $CaF_2$ is performed via a carbon effusion cell in which a $CaF_2$ crystal is heated at ~1300 K, ensuring a total exposure of 1.3 monolayer[38]. The obtained epitaxial surface is then cooled down to ~ 12 K and kept in the preparation chamber, to prepare for molecular adsorption. Sequentially, the $H_2TPP$ molecules are evaporated onto the silicon surface by heating a second effusion (pyrolytic boron nitride, PBN) at ~ 210 °C. During this process, the surface is kept at a low temperature (12 K) using liquid helium cooling of the sample holder. This ensures a gentle landing of the molecules on the substrate and minimizes irreversible surface–molecule interactions. These parameters were carefully optimized beforehand using a quartz balance to achieve homogeneous molecular coverage of approximately 0.2 ML.

The $H_2TPP$ molecule is chosen for this study due to two key characteristics: (i) it exhibits a well-established switching function through the tautomerization of its central hydrogen atoms[56]. (ii) Its lack of a central metallic atom makes it flexible, providing a wide range of vibrational modes. The switching function serves as a signature of the CT in the dyad, while the abundance of vibrational modes facilitates the study of their impact on the entire CT process.

**Formation of the $H_2TPP$ Dyad**

The $H_2TPP$ dyad is fabricated in situ by STM-induced manipulation. It involves either lateral manipulation or local manipulation[57]. We chose to fabricate the dyad composed of two $H_2TPP$-AU molecules that are in a saddle configuration (Figs. S1d) as opposed to the rotated configuration, similar to what is observed with FeTPP/$CaF_2$/Si(100) molecule[40]. The observed meso-N tautomerization process does not result in a 90° switch of the two internal H atoms of the $H_2TPP$ molecule, as often reported on metal surfaces. Our observations reveal a tautomerization resulting from the movement of the two inner H atoms corresponding to a of 45° rotation compared to their initial location. This configuration appears to be energetically preferred due to the slight alternating rotation of the pyrrole groups of $H_2TPP$ molecule, thus reducing the likelihood of a complete 90° rotation of the two hydrogen atoms, as commonly observed in previous work[58].

**STM measurement methods and CT rate estimation**

The electronically induced excitation method of the $(H_2TPP)_2$ dyad is rather straightforward and similar to the procedure used in previous work[59]. The small region at the center of the upper molecule that undergoes a conformational change during the CT process is not located beneath the STM tip and therefore can hardly be reliably detected in the tunneling current trace for statistical analysis. Hence, to excite the lower molecule of the dyad, we use a pre-estimated excitation duration $T_{exc.}$, that provides a probability to induce the CT process of less than unity (i.e. the mean activation time $t_0$ is greater than $T_{exc.}$). The tunnel current during the excitation procedure is then chosen with this criterion as well as controlling the ionization level of the molecule ($\delta z$ = -1.2 Å) and to favor the desired vibrational modes in the dyad. To ease the interpretation of our CT measurements, we first express the molecular change probability as a quantum yield. The measured probability of success $p_s$ to induce CT in the dyad is defined as the ratio between the number of successful excitations and the total number of excitations, for a given excitation current $I_{exc.}$. A quantitative estimate of the CT yield per tunnel electron is given by $Y = e/(I_{exc.} \times t_0)$, where $e$ is the elementary charge and $t_0 = -T_{exc.}/\ln(1 - p_s)$ is extracted from binomial statistics[59]. The electrical work delivered during this characteristic time is $W= V_{exc.} \times I_{exc.} \times t_0$, which defines the energy injected to induce the CT event. Mapping this energy scale onto a frequency scale yields an effective CT rate as $\Gamma( V_{exc.}) = W/h = (V_{exc.} \times I_{exc.} \times t_0)/h$, where h is the Planck constant and $\Gamma( V_{exc.})$ is expressed in $s^{-1}$.

The dI/dV measurements are performed with a double lock-in amplifier setup. One lock-in (master) modulates the voltage bias at a frequency f = 847 Hz with an amplitude of ~ 5 mV. The second lock-in reads this frequency and delivers a signal proportional to the first derivative of the tunnel current when the bias V is tuned. The two dI/dV signals are compared to avoid spurious effects. The ensuing dI/dV measurements are repeated several times, at the same position, and averaged to reduce noise.

After forming the $(H_2TPP)_2$ dyad by lateral manipulation, the STM tip first scans the entire structure to record its topography (−2.2 V, 7 pA) and verify its initial conformation, denoted $(H_2TPP\text{-}AU_\alpha)_2$. The index α indicates that both molecules are in the all-up configuration, i.e., not tautomerized. An excitation pulse is then applied to one of the pyrrole groups ($P_1$–$P_3$), selected randomly. The excitation bias $V_{exc.}$ is varied between −2.0 V and −2.9 V. The pulse duration is first set between 1 and 10 seconds in order to determine the optimal excitation time $T_{exc.}$. After each pulse at a defined $T_{exc.}$, a small area of the upper molecule is scanned to determine whether the electronic excitation has induced tautomerization through a CT process (Supplementary Fig. S10). This highly localized scan minimizes the risk of molecular displacement and prevents dyad destruction. A full topographic scan is avoided because the low diffusion barrier associated with physisorption of the molecule on the $CaF_2$/Si(100) surface makes the system particularly sensitive to tip-induced motion.

Following excitation, two outcomes are possible. If no change is observed in the upper molecule (channel 2a in Supplementary Fig. S10b), the excitation pulse is considered unsuccessful. A small scan of the lower molecule is then performed to confirm the absence of any modification. The dyad is subsequently ready for a new excitation pulse (Supplementary Fig. S10c).

If the upper molecule has changed (channel 2b in Supplementary Fig. S10b), resulting in the formation of the $H_2TPP\text{-}AU_{\beta}^{U}$ conformation, the system must be reset to its initial state. This is achieved by applying a voltage pulse to the left side of the upper molecule (−2.2 V, 10 pA, 2 s; red dot in Fig. S10d). The resetting process is highly efficient and does not disrupt the dyad, allowing reliable recovery of the initial conformation (channel 4 in Supplementary Fig. S10). A new excitation pulse can then be applied.

The transient molecular state $H_2TPP\text{-}AU_{\beta}$ is described in Supplementary Fig. S11. It corresponds to a metastable meso-N tautomerization in which the two central hydrogen atoms are rotated by 45° relative to their initial configuration, while the two lateral pyrrole groups remain slightly tilted upward.

Importantly, the $(H_2TPP\text{-}AU\alpha)_2$ dyad is remarkably stable on the $CaF_2$/Si(100) surface, remaining unchanged for several days in the absence of perturbation. This stability enables the collection of a large number of CT events, providing a statistically robust dataset. Measurements performed on two independent dyads yield sufficient statistics to reduce the error on the mean to approximately 10%.

**Theoretical calculation in gas phase.**

We performed quantum chemical calculations using DFT as implemented in the Gaussian (G16) software package to investigate various conformations of the $H_2TPP$ molecule[60]. In particular, we studied the $H_2$TPP-flat, $H_2$TPP-AU (neutral and cation), $H_2$TPP-UD, $H_2$TPP-AU$_T$ and $H_2$TPP-AU$_\beta$ conformations along with selected cationic forms, in order to determine the total energies of each system and to analyze the energy and spatial distribution of the relevant Kohn-Sham orbitals and the corresponding DOS (Supplementary Figs. S12).

We also extended the analysis to the ($H_2$TPP-AU$_\alpha$)$_2$ dyad, focusing on the spin-polarized Kohn-Sham orbital DOS and the associated vibrational spectra. All DFT calculations were carried out using the CAM-B3LYP functional in combination with the aug-cc-pVDZ basis set. The ultrafine integration grid (int=ultrafine) was employed to ensure high numerical precision and accuracy, particularly for the cationic species. When computing vibrational spectra, we carefully verified the absence of imaginary frequencies to confirm that the optimized structures correspond to local minima rather than saddle points.

**Theoretical calculation of the ($H_2$TPP-AUα)$_2$ dyad on the $CaF_2$/Si(100) surface**

Numerical simulations using density functional theory (DFT) implemented in the SIESTA package were conducted to investigate the adsorption and electronic properties of the ($H_2$TPP)$_2$ dyad adsorbed on a $CaF_2$/Si(100) surface with respect to the orientations observed experimentally (Supplementary Fig. S1)[61,62,63,64]. The structure of the $H_2$TPP molecules placed side by side was optimized using a self-consistent field (SCF) approach. Initial relaxation of the molecules was performed in vacuum (no surface) as described in the section above to minimize strain effects and ensure accurate structural representation in the $H_2$TPP-AU configuration. Subsequently, the molecules were precisely placed on the $CaF_2$/Si(100) slab, which consisted of a periodic volume of $4.64 \times 3.10 \times 4.44$ nm$^3$ and contained 1820 atoms. To model the exchange-correlation interactions, the generalized gradient approximation (GGA) was employed[65], enhanced with van der Waals interactions using the methods of Dion et al. and Román-Pérez et al.[66,67] to account for dispersion forces. The electronic structure calculations utilize a polarized double-$\zeta$ basis set (DZP) and nonlocal norm-conserving pseudopotentials. The simulations featured a mesh cutoff energy of 150 Ry, ensuring high precision in total energy calculations with an accuracy of 1 meV. Brillouin zone integration was performed with a single k-point at the $\Gamma$ center. Geometry relaxations were achieved using the conjugate-gradient method, with forces converged to a threshold of 0.02 eV/Å. For both parallel and tilted configurations, adsorption energy minimization determined the most stable conformations. The Tersoff-Hamann approximation was applied to calculate the local density of states (LDOS) and generate surface-state isodensity maps. Van der Waals interactions, modeled through the dispersion-corrected GGA functional, were essential in describing the adsorption behavior of the $H_2TPP$ molecules; neglecting these interactions led to significantly underestimated adsorption energies[68,69].

## Theoretical calculation of charge transfer rate with Python

The electronic coupling between the two fragments of the $(H_2TPP)_2$ dyad is calculated for a pre-relaxed dimer configuration obtained using Gaussian 16, as described above. All calculations are done in the gas phase. The total CT rate $\Gamma(E)$ (Eq. 1) and the Fragment Orbital Method (FOM) were implemented in a Python code using the quantum-based chemistry package PySCF. The code is available as a Jupyter Notebook on request. Calculations were carried out at the Kohn–Sham DFT (KS-DFT) level of theory, employing the CAM-B3LYP exchange–correlation functional in combination with the cc-pVDZ basis set. This approach enables us to predict and compare the CT yields for different excitation sites, specifically $P_1$, $P_2$, and $P_3$, corresponding to the electronically excited pyrrole subunits described in Fig. 3a.

## ASSOCIATED CONTENT

## Data availability

The data supporting the findings of this study are available in the main text and the Supplementary document. Source data are provided with this paper. Additional data are available from the corresponding authors upon reasonable request.

**Supported information**

Supplementary information available includes: description of the main adsorbed $H_2TPP$ conformations (Fig. S1 and Note 1), and probing of the inner H-atom positions within the pyrrole groups of a single $H_2TPP$ molecule (Fig. S2). Measurements of the $(H_2TPP)_2$ dyad profiles and sizes (Fig. S3), as well as details of the vibrational spectroscopy near the PIR resonance (Fig. S4). Calculated vibrational spectra of isolated $H_2TPP$-AU molecules and $(H_2TPP)_2$ dyads in neutral and cationic states (Fig. S5), together with comparisons between experimental and calculated vibrational energies (Fig. S6). Spatially resolved vibrational modes associated with each pyrrole group (P1–P3) in the $(H_2TPP)_2$ dyad (Figs. S7–S9), and detailed description of the excitation procedure (Fig. S10). Structural evolution of the dyad during the CT-induced conformational change (Fig. S11). Calculated energy differences between relevant $H_2TPP$ conformations and transient states (Fig. S12 and Notes 2–3). Details of the CT rate extraction and fitting procedures (Fig. S13 and Note 4), including analysis of the reaction coordinates and their relationship to the reorganization energy (Notes 5–6). Comparison of spin-polarized orbital energies within the dyad (Fig. S14 and Note 7), and three-dimensional representation of the CT rate within the MLJ framework (Fig. S15 and Note 8).

**AUTHORS INFORMATION**

Jessica Martinez, ORCID N° 0000-0002-9577-3449, jam1042@scarletmail.rutgers.edu

Prof. Michele Pavanello, ORCID N° 0000-0001-8294-7481, m.pavanello@rutgers.edu

Dr Eric Duverger, ORCID N° 0000-0002-7777-8561, eric.duverger@univ-fcomte.fr

Dr. Damien RIEDEL, ORCID N° 0000-0002-7656-5409, damien.riedel@universite-paris-saclay.fr

Corresponding author: Dr. Damien RIEDEL, damien.riedel@universite-paris-saclay.fr,

web site: http://mnd-sciences.com

**Competing interests**

The authors declare no competing financial interests.

## ACKNOWLEDGMENTS

This work is supported by the ANR through the CHACRA project (ANR-18-CE30-0001). DR would like to thank *the federation lumière matière* (LUMAT) for the access to the MESOLUM calculation cluster. E.D. wishes to thank *the Mesocentre de calcul* of the UMLP University and the Communauté d'Agglomeration du Pays de Montbeliard (convention PMA-UFC). All the authors discussed the results and contributed to the manuscript writing. D.R. supervised the project.

## Author contribution

D.R. wrote the article, performed the STM experimental measurements and analysis of the data as well as numerical simulations with Gaussian 09 software. D.R. developed the combined Antoniewicz-Marcus model and ensuing calculations. E.D. has done the DFT simulations using the SIESTA code of the $H_2TPP$ dyad when adsorbed on the $CaF_2$/Si(100) surface. J.M. and M.P. developed the Python code within the Pyscf package to estimate the CT rate when the $H_2TPP$ dyad is in gas phase. All the authors have proof read the entire article.

## Figures Captions:

**Figure 1 | CT platform at the nanoscale and $H_2TPP$ conformations on the $CaF_2$/Si(100) surface. a,** Ball-and-stick representation illustrating the principle of the nanoscale CT platform based on an $(H_2TPP)_2$ molecular dyad. **b.** and **c.** Ball-and-stick representations of the two $H_2TPP$ molecules in the $H_2TPP$-UD (**b**) and $H_2TPP$-AU (**c**) conformations, respectively. **d.** and **e.** Top-view spatial distributions of the density

of states (DOS) for the $H_2$TPP-UD (d) and $H_2$TPP-AU (e) conformations. In both cases the H atoms are oriented vertically.

**Figure 2 | Electronic and vibrational structure of the adsorbed $H_2$TPP-AU molecule and dyad. a.** dI/dV spectrum acquired at several points (see inset) on a single $H_2$TPP-AU molecule. **b.** Calculated Kohn-Sham orbitals of the $H_2$TPP-AU conformation (gas phase) with indications of the HOMO-LUMO gap energy. The right panel in (**b**) shows the calculated projected density of states (PDOS) on the pyrrole and phenyl groups of the $H_2$TPP-AU molecule when adsorbed on the $CaF_2$/Si(100) surface, for comparison. **c.** STM topography (53 × 45 Å², -1.9 V, 7 pA) of a $(H_2TPP\text{-}AU)_2$ dyad during its formation, shown prior to the final lateral manipulation (dark-yellow arrow). Side view: STM topography of the $(H_2TPP\text{-}Au)_2$ dyad after its formation (17 × 37 Å², −1.9 V, 7 pA). **d.** dI/dV spectra acquired at three different tip heights.: $\delta z$ = 1.0 Å, $\delta z$ = 1.5 Å, $\delta z$ = 1.8 Å at $P_1$. The presented dI/dV curves are consistent across all probed locations $P_1$, $P_2$, and $P_3$, as indicated in the right panel of **c** on the $(H_2TPP\text{-}Au)_2$ dyad. **e.** Measured dI/dV vibrational spectra at point $P_1$ extracted from the corresponding dI/dV curves, using the PIR energy (-1.78 eV) as a reference. Representative vibrational modes are indicated ($cm^{-1}$), by red vertical arrows and grouped by energy range.

**Figure 3 | Tautomerization, CT-rate measurement and modeling excitation in the $(H_2TPP\text{-}AU)_2$ dyad. a.** and **b.** STM topography (17 × 37 Å², -1.9 V, 7 pA) of the $(H_2TPP\text{-}AU)_2$ dyad before (**a**) and after (**b**) the CT process inducing the tautomerization detected in the upper molecule (see orange dotted square). The notation $H_2TPP\text{-}AU_\alpha^U$ and $H_2TPP\text{-}AU_\alpha^L$ refers for the upper 'U' and lower 'L' molecule in the dyad as oriented in the STM image. The indices α, β indicate if the molecule has switched (β) or not (α) to its tautomer conformation (i.e. $H_2TPP\text{-}AU_\beta^U$). **c.** Measured CT rate (blue circles) as a function of the bias voltage applied at each excitation point $P_1$ - $P_3$. The error bars are indicated in light blue. The red dotted curves are the resulting numerical fit based on the MLJ model described in equation (2) (see main text). **d.** and **e.** Calculated CT rate in the dyad based on the Fermi Golden Rule model described in equation (1)

when the dyad is in its neutral state (**d**) or cationic state (**e**). **f.** Simplified band diagram of the STM junction illustrating the excitation process that generates a local cationic state in the dyad beneath the STM tip together with the associated vibrational modes. The amount of energy $qV_{exc.}$ represents the excitation energy at the applied bias voltage, while $\gamma qV_{exc.}$ denotes the fraction of this energy that is used to vibrationally populate the dyad above the ionization potential.

**Table 1 | Physical parameters extracted from the measured data at the three excitation points $P_1$ - $P_3$.** For each excitation point $P_1$, $P_2$, and $P_3$, the values of $\Delta G_{@max}$, $\Gamma_{max}$, FWHM, $\hbar\omega_e$, $\lambda_s$, $S_e$, , $\lambda_i$, T (and $k_BT$) as well as $|V_{ij}|^2$, $G_0$ , and $\Delta G_0$ are estimated for each mode n. These parameters represent respectively: the fitted peak position, the maximum CT rate, full width at half maximum of each CT-rate peak, the vibrational wavenumber, the reorganization energy, the Huang-Rhys factor, the ensuing internal reorganization energy, the fitted temperature value, the electronic coupling, the baseline fitting parameter, and the Gibbs free energy of the cation state, respectively. The estimated errors on the fitted values are provided in Supplementary Note 4.

**Figure 4 | Step-by-step evolution of the $(H_2TPP\text{-}AU)_2$ dyad dynamics along two reaction coordinates. a.** to **d.** Series of schematic representations illustrating the successive steps of the CT process in the $(H_2TPP\text{-}AU)_2$ dyad introducing the relevant parameters used in the MLJ-Antoniewicz model: The two $H_2TPP$-AU molecules named $M_1$ and $M_2$ are represented by oblate spheroids which color changes with their electronic states: **a** before any excitation, **b** after the extraction of an electron in $M_1$, **c** after the charge inversion in the dyad from $M_2$ to $M_1$, **d** after the tautomerization and neutralization of the upper fragment in the dyad ($M^T_2$). The image charge is indicated at the surface by a (-) sign when needed. **e.** to **g.** Morse potential curves illustrating the three steps of the CT process induced in the $(H_2TPP\text{-}AU)_2$ dyad within the Antoniewicz model along the $h$ coordinate axis: **e** ionization step during the formation of a local cation. **f.** Charge inversion in the dyad. **g.** Tautomerization and neutralization of the dyad (surface). **h.** to **j.** Schematic harmonic potential curves illustrating the evolution of the system within the Marcus model

along the $d$ coordinate axis: **h** after its ionization when $d = d_c$, **i** during the charge inversion process, **j** during the tautomerization and neutralization of the dyad.

**Figure 5 | 3D schematic of the dyad's electronic and vibronic states with calculated evolution of the reorganization energy $\lambda_s$.** **a.** Qualitative 3D representation (truncated) of the potential energy surfaces (PES) derived from a combined Marcus-Antoniewicz model during the first step of the ionization of the excited molecule. The blue and orange PES represent the neutral and cationic $(H_2TPP\text{-}AU)_2$ dyad states, respectively. The white dotted parabolic and Morse potential profiles are indicated to clarify the PES construction. **b.** Qualitative 3D representation of the PES following an electronic excitation, after the system has evolved quantum mechanically, inducing variations in $\Delta h$ and $\Delta d$ which serve as precursors to the second step of the CT process (charge inversion) in the $(H_2TPP\text{-}AU)_2$ dyad. **c.** Qualitative 3D representation of the PES shown for the neutral state (blue), the cationic state of the first $H_2TPP\text{-}AU$ fragment (orange), and the cationic state of the second $H_2TPP\text{-}AU$ fragment (green) during charge inversion when the voltage reaches PIR + $\gamma qV_{exc}$ to match vibrational modes $\nu_m$ and $\nu_n$ that trigger the ad-hoc Frank-Condon transition. $\Delta h_T$ and $\Delta d_{CT}$ variations are indicated for clarity. (d) Calculated variations of $\lambda_s(h_0)$ when $h_0$ varies from 0 to 14 Å, for R = 5.6 Å, $d_0$ = 14.6 Å and $\lambda_i$ = 0.3.